\documentclass[final,5p,times,twocolumn]{elsarticle}
\usepackage{amssymb}
 \usepackage{amsthm}
 \usepackage{amsmath}
\usepackage{booktabs}
\usepackage{siunitx}
\usepackage{cleveref}
\usepackage[algoruled,boxed,lined]{algorithm2e}

\usepackage{epstopdf}

\journal{}

\begin{document}

\begin{frontmatter}



\title{Investigating the Thixotropic Behaviour of Tremie Concrete Using the Slump-flow Test and the Material Point Method}


\author[label1]{Christopher Wilkes}
\author[label2]{Krishna Kumar}
\author[label1]{Giovanna Biscontin}

\address[label1]{Department of Engineering, University of Cambridge, Cambridge CB2 1PZ, United Kingdom}
\address[label2]{Dept of Civil Architectural and Environmental Engineering, Cockrell School of Engineering, University of Texas at Austin, Texas, USA}

\begin{abstract}

This paper presents a new thixotropic model integrating the Papanastasiou-Bingham model with thixotropy equations to simulate the flow behaviour of Tremie Concrete in the Material Point Method framework. This study investigates the effect of thixotropy on the rheological behaviour of fresh concrete by comparing field measurements with numerical simulations. The comparison yields new insights into a critical and often overlooked behaviour of concrete. A parametric study is performed to understand the effect of model parameters and rest-time on the shear stress response of fresh concrete. The Material Point Method with the Papanastasiou-Bingham model reproduces Slump-flow measurements observed in the field. The novel model revealed a decline in concrete workability during the Slump-flow test after a period of rest due to thixotropy, which the physical version of the test fails to capture. This reduction in workability significantly affects the flow behaviour and the effective use of fresh concrete in construction operation. 
\end{abstract}


\end{frontmatter}



\section{Introduction} 
\label{Introduction}
Tremie Concrete is widely used to construct bored piles and diaphragm walls for its superior post-hydration compressive strength and high level of workability. The term workability is often used interchangeably with consistence; however, both terms are useful descriptors of the relative mobility or the ability of fresh concrete to flow~\citep{EFFC2}. The requirements for the use of Tremie Concrete in deep foundations includes its ability to flow around obstacles like a congested reinforcement cage and freely through a pipe, thereby, needing a high degree of consistence. When left undisturbed, tremie concrete stiffness will progressively increase during the dormant period of the hydration reaction~\citep{roussel06, roussel2012origins}, leading to a reduction in the ease with which it can flow. Fortunately, the loss of mobility due to the increase in stiffness can be recovered with an application of stress, so long as the applied stress is significant enough to break down the hydration byproducts causing the elevated strength \citep{roussel2012origins}. The temporary reduction in concrete mobility that can be alleviated by stress application is referred to as thixotropy.

Thixotropy poses a severe issue in the construction industry as changes to flow behaviour can hinder the ability of new concrete to mix with the existing concrete, thereby creating weak interface planes~\citep{roussel06}. Furthermore, thixotropic changes in concrete could prevent concrete from flowing freely around reinforcement bars during long pile casting operations~\citep{thorp}.

Increasingly, the suitability of a concrete mix design is examined through numerical analysis of the construction operation~\citep{EFFC2,sofcf,claud_thesis,Vasilic_Gram_Wallevik_2019, roussel20}. However, the most popular approach of modelling uses Computational Fluid Dynamics based numerical methods which encounter difficulty when simulating thixotropic behaviour due to the Eulerian frame of reference. The Material Point Method (MPM)~\citep{sulsky94,sulsky95} is an emerging numerical method capable of simulating large deformation and time-dependent material allowing for the simulation of thixotropic concrete.

The objective of this paper is to develop a comprehensive thixotropic model in the Material Point Method (MPM) framework, to accurately capture the history-dependent nature of the flow behaviour of Tremie Concrete. This paper provides unique insights into the history-dependent nature of Tremie Concrete and the associated flow response by exploring the challenges associated with testing and simulating thixotropic concrete.

\section{Flow Behaviour} 
\subsection{Slump-flow test}

\begin{figure}
\centering
\includegraphics[width=\linewidth]{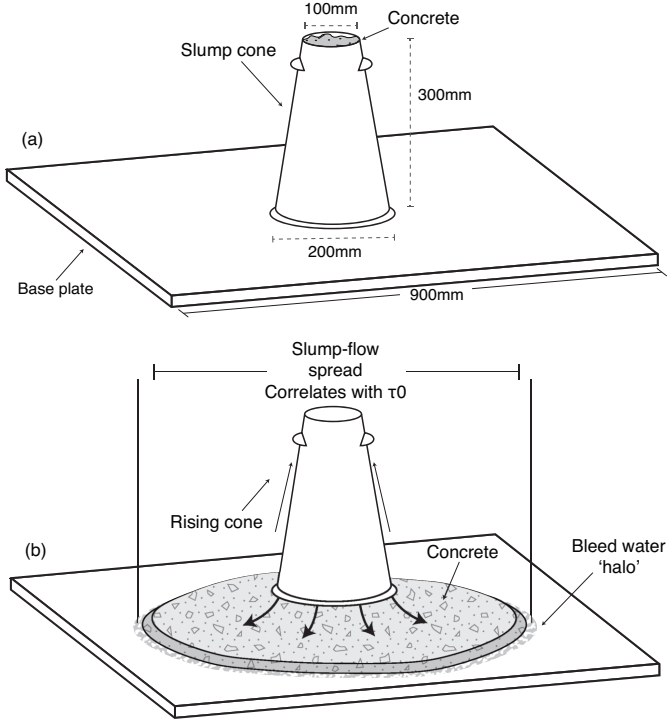}
\caption{\label{fig:slumptest}a) Slump-flow test apparatus in accordance with BS EN 12350-8:2019. b) Representation of a typical Slump-Flow spread (SF).}
\end{figure}

The Slump-flow test~\citep{BS123508} (\cref{fig:slumptest}), is an empirical test that quantifies a concrete's flow behaviour by measuring Slump flow spread (SF) - the distance concrete spreads before stopping when emptied from a rising cone~\citep{Sperwall}. Even in its simplicity, the SF test is able to capture a snapshot of concrete flow behaviour at a single point in time, however, the test has drawn criticism for its inability to observe or provide insight on the behaviour of concrete following a period of rest~\citep{krankel,dfi_feys}. The SF of a Tremie Concrete is typically in the range of \SI{500}{\mm} to \SI{600}{\mm}~\citep{Sperwall}, although, in practice, concretes with a SF as low as \SI{450}{\mm} or as high as \SI{650}{\mm} are also considered acceptable~\cite{EFFC2}. Recent analysis of concrete SFs~\citep{wallevik06,roussel50,saak04} found the Bingham rheological model is able to accurately describe the flow behaviour of fresh concrete~\citep{krankel,dfi_feys}.

A material's rheology provides a quantifiable description of its flow properties. The Bingham rheological model~\citep{tattersall83} is represented as:

\begin{equation}
\tau = \mu\dot{\gamma}+\tau_0\,,
\label{eq:bingham}
\end{equation}

\noindent where $\tau$ represents the shear stress, $\tau_0$ the yield stress, $\mu$ the plastic viscosity and $\dot{\gamma}$ the shear rate. The yield stress, $\tau_0$, represents the initiation force needed to commence flow but both the yield stress and plastic viscosity $\mu$ control the flow behaviour. In the Bingham model, materials experiencing stresses below a critical yield stress behave akin to rigid bodies with viscoelastic effects, whilst stresses that exceed the yield criteria cause the material to flow and behave as a viscous liquid~\citep{rheologybook}.

\begin{figure}
\includegraphics[width=\linewidth]{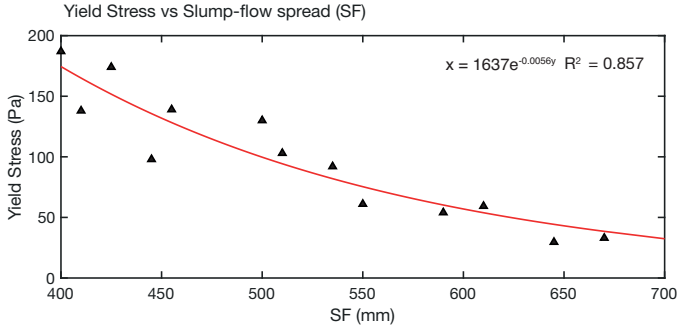}
\caption{\label{fig:t0_sf}Experimental correlation between dynamic yield stress and slump flow (redrawn after~\citep{krankel, dfi_feys}).}
\end{figure}
% Write a sentence explaining experimental correlations are needed to determine rheological properties.

SF measurements and rheological models such as the Bingham model, are fundamental to validating numerical methods. Rheological measurements obtained from a rheometer can be input directly into a numerical model and the associated flow behaviour, represented by an SF, observed and compared with expectations. Thus, SF results paired with rheological measurements not only provide a way to validate a numerical model, but also a pathway to overcome the limitations of verifying the construction operation.

~\Cref{fig:t0_sf} shows the experimental correlation between yield stress derived from rheometer and SF. The SF is observed to increase exponentially as the yield stress decreases~\citep{wallevik06,roussel50,saak04}. When the shear stresses generated during the slump flow test are lower than the yield stress of the concrete, the flow stops~\citep{sofcf}. Thus, any change in yield stress will inherently influence the distance at which the flow stops, creating the yield stress and SF relationship demonstrated in~\Cref{fig:t0_sf}.

Plastic viscosity also impacts the flowability, however, its influence on the SF spread is considered less significant than that of the yield stress as no strong correlations have been established with the stopping distance of the concrete spread. The plastic viscosity does effect the rate at which concrete spreads~\citep{rheologybook,feys2016concrete,sofcf,krankel,dfi_feys}, so an increase in plastic viscosity increases the time taken for concrete to stop spreading, referred to as SF-Time by~\citet{EFFC2}. However, it is worth noting that~\citet{Deeb2014} and~\citet{sofcf} showed that the speed at which the cone is raised influences the spreading rate, without changing the final SF~\citep{nielsson2003rheological, tanigawa1989analytical}. The relation between SF-Time and plastic viscosity may depend on the variance between different operators raising the cone at different speeds, within the Slump-Flow test standard~\citep{BS123508}. To validate a rheological model, it is important to compare against the final run-out distance rather than the SF time. The SF time should not be used solely to confirm if a numerical model is reliable.

Where the input values for a simulation are taken from field measurements, the accuracy of a Slump-Flow simulation can be judged largely by considering the level of agreement between the simulated SF and field SF, as the yield stress/ SF relationship is subject to less controversy and user-variance. Although the yield stress/SF correlation is better than that of plastic viscosity/ SF-time, there is still variance in the behaviour of concrete as the yield stress changes at rest due to thixotropy; which cannot be easily captured by the Slump-Flow test. Thus, the variability in the yield stress and the impact thixotropy has on the Slump-flow test must be evaluated in order to use SF as a validation method for time-dependent concrete flow behaviour. 

\subsection{Thixotropy} 
\label{section:thix}
The yield stress of a cement-based material like concrete originates from the micro-structure of cement particle-particle networks through colloidal interactions or direct contact between cement particles~\citep{qian2018distinguishing}. This micro-structure can sustain a certain amount of stress before it is broken down and starts to flow. In the Bingham model of concrete, the resistance to flow due to the micro-structure is defined as the yield stress~\citep{roussel06}. At rest, this micro-structure will continue to build up due to colloidal flocculation and formation of calcium silicate hydrate (CSH) bridges between particles as a result of partial hydration of cement~\citep{roussel2012origins}, resulting in increased yield stress~\citep{lecompte2017non}, leading to thixotropy. When sufficient shear force is applied, the micro-structure will begin to break down, causing the elevated yield stress to dissipate to a constant yield stress upon which no further breakdown of the structure is possible. The temporary increase in yield stress after a period of rest is referred to as the `\textit{static}' yield stress, while the constant yield stress observed once the elevated level has dissipated is referred to as the `\textit{dynamic}' yield stress~\citep{qian2018distinguishing}.

\citet{roussel06} proposed a series of equations to characterise the influence of thixotropy on yield stress. Incorporating a flocculation state $\lambda$ in \cref{eq:bingham} gives:

\begin{equation}
\tau = \mu\dot{\gamma}+(1+\lambda)\tau_0\,,
\label{eq:thix}
\end{equation}
\begin{equation}
\frac{\partial \lambda}{\partial t}=\frac{A_{thix}}{\tau_0}-\alpha\lambda\dot{\gamma}\,,
\label{eq:thixdt}
\end{equation}
where $A_{thix}$ is the restructuring rate at rest and $\alpha$ is the destruction parameter (typically of the order $0.01$)~\citep{roussel07}. Considering the rate of build-up of the micro-structure, $A_{thix}$, is relatively slow compared to the rate of destructuration, as reported by~\citep{papo1988thixotropic, roussel06}, \cref{eq:thixdt} becomes:
\begin{equation}
\frac{\partial \lambda}{\partial t}=-\alpha\lambda\dot{\gamma} \,.
\label{eq:thixdt_const}
\end{equation}
integrating~\cref{eq:thixdt} with respect to time, we get:
\begin{equation}
\lambda = \lambda_0e^{-\alpha\dot{\gamma}t}\,,
\label{eq:thix_int}
\end{equation}
where, $t$ is time spent at a particular shear rate and $\lambda_0$ is the original flocculation state. The three thixotropic parameters ($\lambda, \alpha$ and $A_{thix}$) can be computed by measuring the torque required to shear a rested concrete.

\begin{figure}
\includegraphics[width=\linewidth]{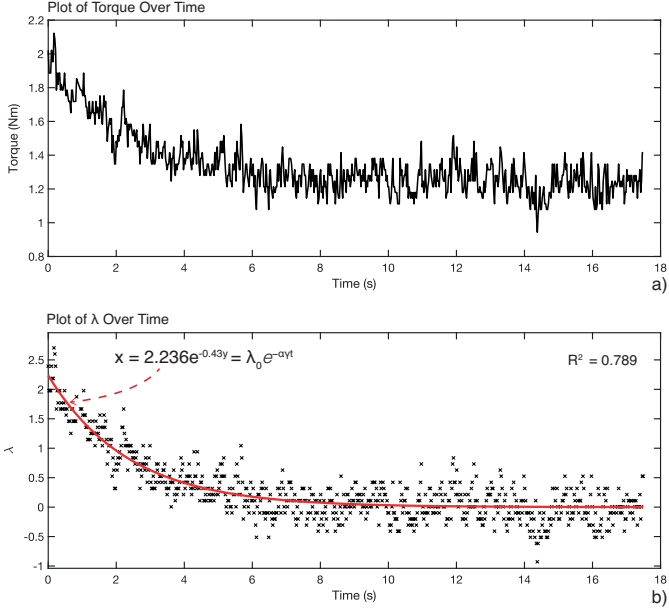}
\caption{\label{fig:torque} a) Evolution of torque with time, measured by a rotating vane submerged in Tremie Concrete, with a shear rate of \SI{12.43}{\per\second}. Torque decaying over time to a constant average. b) Evolution of $\lambda$ over time with a fitted exponential curve. Data provided by internal communications with ARUP.}
\end{figure}

\Cref{fig:torque}a shows a gradual decay of observed torque with time for a Tremie Concrete using an ICAR~\citep{ICAR} rheometer at a site within the United Kingdom. Coaxial rheometers such as the ICAR~\citep{ICAR} use a rotating vane submerged in a cylinder filled with concrete to measure torque. Bingham parameters are then derived from the torque and rotational velocity using the Reiner-Riwlin equations~\citep{wallevik2015avoiding,feys2018rhe}. Incidentally, the observed torque, $T$, will gradually decay from an elevated initial measurement to a stable value. In~\cref{fig:torque}a, torque reduces from approximately \SI{2}{\newton\meter} to a stable value of \SI{1.25}{\newton\meter} after shearing (rotation of the vane) for approximately \SI{10}{\second}. The reduction from $2$ to \SI{1.25}{\newton\meter}, assuming a constant rotational velocity of the vane, indicates the transition from static yield stress to dynamic yield stress. The thixotropic parameters of the Bingham model can be calculated using the apparent shear stress, defined as a function of the observed torque $T$, radius $r$ and height $h$ of the vane. The apparent shear stress is measured in a rheometer as:

\begin{equation}
\tau_{apparent} = \frac{T}{2 \pi r^2h} = \mu\dot{\gamma}+(1+\lambda)\tau_0\,.
\label{eq:torque}
\end{equation}

From \cref{eq:torque}, $\tau_{apparent}$ can be used to calculate a flocculation state $\lambda$ for a given apparent shear stress. Thus, by taking $\tau_{apparent}$ at different time intervals and using it to calculate the corresponding flocculation state of the concrete, a $\lambda$/time plot can be generated as shown in \cref{fig:torque}b. At the maximum $\lambda$, the elevated static yield stress can be calculated as:

\begin{equation}
(1+\lambda)\tau_{0_{dynamic}} = \tau_{0_{static}}\,.
\label{eq:static}
\end{equation}

Despite noisy data measured in a rheometer, an exponential best-fit curve can be applied to the $\lambda$/time plot (\cref{fig:torque}b) to generate an approximation of $\alpha$. The coefficient of the exponent, $0.43$, from \cref{fig:torque}b is divided by the known constant shear rate, \SI{12.43}{\per\second}, as per \cref{eq:thix_int}, to yield an approximate $\alpha$ of $0.0358$.

In the present study, an $\alpha$ of $0.01$ is used for all concretes as the raw rheometer data for the examined concretes is unavailable. However, this value represents an $\alpha$ supported by \citet{roussel06}. 

The remaining thixotropic variable, $A_{thix}$, which represents the rate at which yield stress increases over time (\SI{}{\pascal\per\second}), can be calculated using the static yield stress in \cref{eq:static}:

\begin{equation}
A_{thix} = \left(\tau_{0_{static}}-\tau_{0_{dynamic}}\right)/rt\,.
\label{eq:athix}
\end{equation}

\begin{figure}
\includegraphics[width=\linewidth]{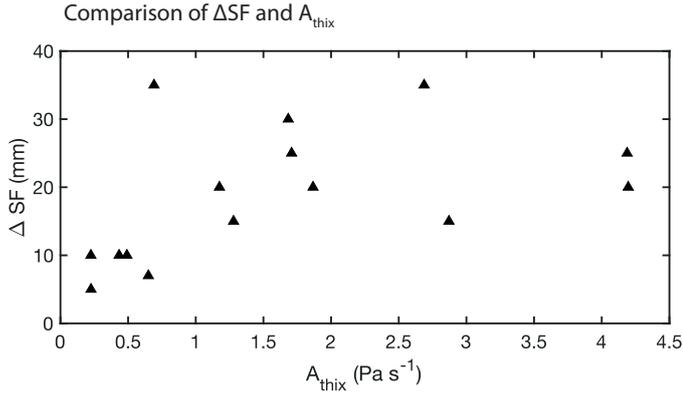}
\caption{\label{fig:thix_comp} Comparison of change in SF at \SI{240}{\second} and \SI{0}{\second} ($\Delta$SF) with $A_{thix}$. Created using data from~\citep{krankel, dfi_feys}.}
\end{figure}

$A_{thix}$ is used as a single value representing the thixotropic propensity of concrete, where a value of $A_{thix} > 0.5$ is considered to be highly thixotropic~\citep{roussel06}. $A_{thix}$ values greater than $3.0$ for some Tremie Concretes were reported by \citep{krankel,dfi_feys} for concrete with a rest time of \SI{240}{\second}. This represents a large increase in yield stress over a relatively short period. It is therefore reasonable to expect a significant decline in SF after a period of rest for highly thixotropic concrete, given such a large increase in yield stress.
The difference in Slump-flow before and after a period of rest is written as $\Delta$SF, where:

\begin{equation}
\Delta SF = SF_0 - SF_{240}\,.
\label{eq:deltasf}
\end{equation}

with SF$_0$ and SF$_{240}$ representing the SF measured following \SI{0}{\second} of rest and \SI{240}{\second} of rest respectively.

The comparison of slumps before and after \SI{240}{\second} for different concrete mixes, shown in \cref{fig:thix_comp}, indicates only small changes in SF despite large $A_{thix}$ values~\citep{krankel}, although a general pattern of increasing $\Delta$SF with $A_{thix}$ is observed, no  confident trend can be identified. Upon further inspection, a mix with an $A_{thix}$ of $2.87$ resting for $240$ seconds experienced a reduction in SF of only \SI{15}{\mm} despite an increase in yield stress of \SI{688}{\pascal} from an initial value of \SI{130}{\pascal}. Given the relationship between yield stress and SF shown in \cref{fig:t0_sf}, a high value of yield stress should result in a significantly reduced SF that is no longer within the acceptability limits for a Slump-flow test, clearly that did not occur in this case. This lack of SF reduction can be explained partially by the underlying relationship between yield stress and plastic viscosity created by mix designers to maintain stability at low yield stresses or low viscosities. To maintain stability, the yield stress of a low viscosity concrete would be increased, and vice versa. These alterations are conducted by altering the mix design, modifying fines or water content, or the use of additives. Thus, an elevated yield stress, relatively low viscosity concrete will not have the same SF/Yield stress relationship as in~\cref{fig:t0_sf}. Nevertheless, given the significant effect yield stress has on SF, a decrease in SF of at least \SI{100}{\mm} would not be unexpected for an increase in yield stress as dramatic as \SI{688}{\pascal}.

The unexpectedly small $\Delta$SF values shown in \cref{fig:thix_comp} can be further explained by examining the experimental procedure used to gather the data. During the \SI{240}{\second} rest period, the concrete remains undisturbed in the cone on a base-plate. As the cone is not sealed on the base plate, bleed water from the concrete can accumulate at the base and seep out. The test standard BS EN 12350-8:2019~\citep{BS123508} specifies a maximum wait time of \SI{30}{\second} to avoid potential issues like dewatering, as any water at the base can reduce friction between the concrete and the base plate, leading to a larger SF. \Citet{krankel} also reported a \SI{30}{\mm} variation in SF within the allowable resting time, indicating potentially large inaccuracies. Some concretes in \cref{fig:thix_comp} saw a $100\%$ increase in yield stress in just \SI{30}{\second} of rest. Considering that BS-EN 12350-8:2019~\citep{BS123508} allows up-to \SI{30}{\second} rest before the cone is raised, it is conceivable that the concrete has already accrued some degree of thixotropy in the regular test that the SF$_0$ figures represent. 

Hence, obtaining any meaningful thixotropy effects in a physical Slump-flow test can become highly challenging~\citep{krankel, dfi_feys, roussel2007a}. If the thixotropic behaviour of concrete is not correctly understood, issues of restricted concrete flow behaviour during the construction process could occur. Hence, the measurements of thixotropy can be of great importance in practice. Numerical models can minimise the variability of physical tests by allowing for greater control over all variables, creating an ideal testing situation, thereby offering unique insights into concrete flow behaviour by accurately considering the influences of thixotropy which are otherwise obscured by physical testing.

\section{Materials and Methods} 
\subsection{Numerical Model} 
The behaviour of fresh concrete is typically modelled as a continuum using Lagrangian methods~\citep{Deeb2014, Mechtcherine2014} or Eulerian Computational Fluid Dynamics (CFD)~\citep{EFFC2, roussel50, sofcf, roussel07, vasilic2016flow, claud_thesis}.~\Citet{Schryver2018} modified OpenFOAM\textsuperscript{\textregistered} (Open Source Field Operation and Manipulation), a popular CFD solver, to incorporate time-dependent thixotropic concrete behaviour. However, at the time of writing the method remains unvalidated. A comprehensive review of numerical methods used in the modelling of concrete flow is presented by~\citet{sofcf} with more recent model reviews presented by~\citep{Vasilic_Gram_Wallevik_2019,roussel20}. Continuum methods are suitable for large scale problems but may struggle simulating less workable concrete where inter-granular relationships become more important. Discrete particle methods~\citep{Mechtcherine2014} or Smooth Particle Hydrodynamics~\citep{Deeb2014} methods that represent concrete as individual particles capture the micro-scale behaviour accurately, however, these methods are computationally intensive and cannot always be scaled to the full field-scale. Hence, there is a need for continuum-based numerical methods capable of incorporating micro-structural laws in describing the macroscopic rheological response.

MPM~\citep{sulsky94,sulsky95} offers a new hybrid Eulerian-Lagrangian approach that combines the Lagrangian benefits of point tracking with the Eulerian background grid for modelling large deformation problems with history-dependent materials~\citep{Krishna,Samila}. The MPM uses a background grid on which Lagrangian material points could traverse, thereby avoiding the need for frequent re meshing to avoid mesh distortion~\citep{kenichi_geo}. MPM has been used to successfully model large deformation \citep{Krishna}, interaction \citep{rohe2017}, and history dependent materials \cite{bandara15} with a high degree of accuracy. In this study, MPM is used for the first time to model concrete flow behaviour.

\begin{figure}
\centering
\includegraphics[width=\columnwidth]{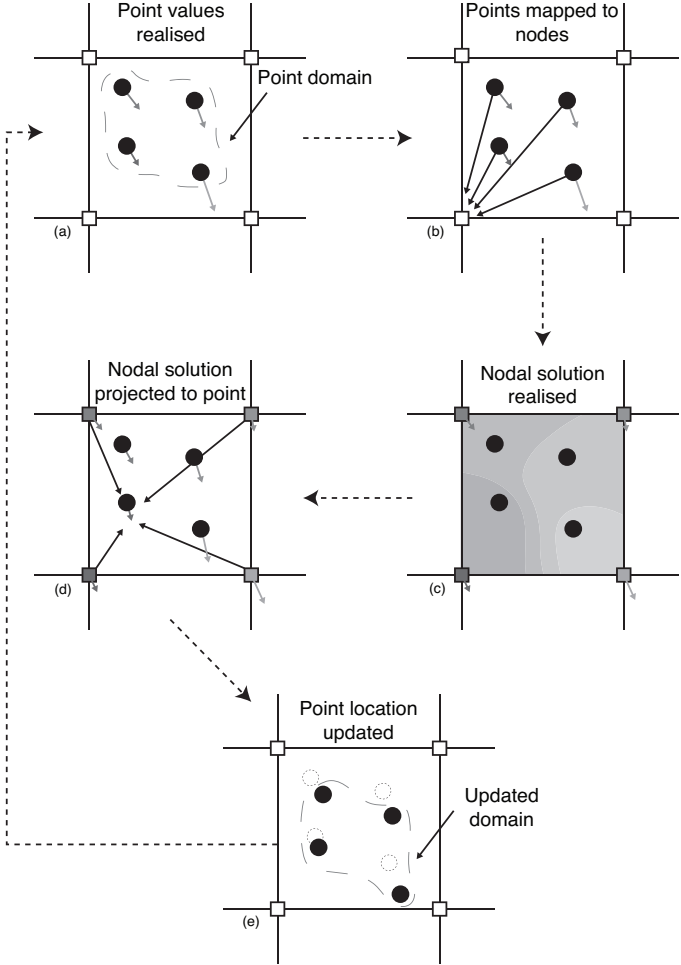}
\caption{\label{fig:nodal}Schematic diagram of the material point algorithm, arrows represent material point properties (velocity, mass, etc.) a) Point values are initialised b) Point values are mapped to nodes c) The equations of momentum are solved on the nodes d) Updated Nodal values are projected back to points e) Point locations and states are updated.}
\end{figure}

The MPM involves taking a continuum body and discretising it into a finite set of material points, the tracking of properties at these points represents the Lagrangian portion of the hybrid approach. The Eulerian part of MPM comes from a background grid on which the momentum equation is solved. This background grid encompasses the entire domain of the problem. Typically, the MPM Eulerian background grid is a Cartesian domain of rectilinear elements. Each element is defined by nodes, usually at the vertices of the element unless otherwise specified. The material points carry the material properties, such as density, stresses, shear strength and viscosity, and are specified as an input. \Cref{fig:nodal} presents a schematic diagram of the MPM solution process. The shape-functions, which are derived from the position of the material points within the element, are used to map material point properties to the nodes, \cref{fig:nodal}a-b,~\citep{Krishna}. The equation of motion is solved at the background mesh to find the current acceleration, with the element integration based on the material point locations {~\cite{wang16}}, \cref{fig:nodal}c. The grid nodal acceleration and velocity values are mapped back to the points to update their velocities and positions using the same shape functions as before~\citep{zhang16}, \cref{fig:nodal}d. The stresses are computed at each material point using a constitutive model. Finally, the locations of the material points are updated based on the mapped velocity from the node as shown in~\cref{fig:nodal}e. The mesh is reset at the end of each iteration, therefore, no permanent information is stored at the nodes.

\subsubsection{Governing Equations}
Neglecting the thermal effects, the governing equations for the MPM are derived from the standard conservation equations for mass and momentum, \cref{eq:mass} and~\cref{eq:momentum} respectively. Derivations presented here are based on work conducted by~\citep{sulsky95,Samila,Krishna,zhang16,Chen2002}.

\begin{equation}
\frac{d\rho}{dt} + \rho\nabla\cdot \mathbf{v} = 0\,,
\label{eq:mass}
\end{equation}
\begin{equation}
\rho a = \nabla\cdot\mathbf{\sigma} + \rho b\,.
\label{eq:momentum}
\end{equation}

\noindent where $\rho(\mathbf{x},t)$ represents the mass density of a point at location $\mathbf{x}$ and time $t$, $v(\mathbf{x},t)$ represents the velocity, $a(\mathbf{x},t)$ the acceleration, $\sigma(\mathbf{x},t)$ the Cauchy's stress tensor and $b(\mathbf{x},t)$ the body force at the point. $\nabla\cdot$ and $\nabla$ represent the divergence and gradient operators, respectively.

A continuum body is discretised into a finite number of material points $n_p$ and let $\mathbf{x}_p^t$ represent the current position of point $p$ at time $t$ where $p=(1,2,...n_p)$. At any given time $t$, each point will have an associated mass $m_p^t$, density $\rho_p^t$, velocity $\mathbf{v}_p^t$, and stress and strain $\mathbf{\sigma}_p^t$ and $\epsilon_p^t$, respectively, providing a Lagrangian description of the body. In addition the material points also track parameters required by the chosen constitutive model. To obtain the discrete form of the equation, \cref{eq:momentum} is multiplied by weight function $w$ and integrated over boundary $\Omega$ to give

\begin{equation}
\int_\Omega \rho w \cdot a\:d\Omega = \int_\Omega w\nabla\cdot\sigma \:d\Omega + \int_\Omega \rho w b \:d\Omega \,.
\label{eq:weightfunction1}
\end{equation}

After integration by parts, the application of the divergence theorem and the inclusion of Neumann and Dirichlet boundary conditions, the weak form of the equation of motion is given by

\begin{equation}
\int_\Omega \rho w \cdot a\:d\Omega = \int_{\partial\Omega_t} w\cdot\tau\: dS - \int_\Omega \nabla w\cdot\rho\sigma^s \:d\Omega + \int_\Omega \rho w b \:d\Omega \,.
\label{eq:weakform}
\end{equation}

Where $\tau$ represents surface traction, $\rho\sigma^s$ represents the specific stress ($\sigma^s = \sigma / \rho$) necessary for the derivation of discrete equations, and $dS$ and $\:d\Omega$ represent surface and volume differential respectively. \Cref{eq:mass} is automatically satisfied as material points have a fixed mass at all times. As the material is discretised into points, the mass density can now be written as

\begin{equation}
\rho(\mathbf{x},t) = \sum_{p=1}^{n_p} m_p \delta(\mathbf{x-x}_p^t)
\label{eq:dirac}
\end{equation}

Where $\delta$ represents the Dirac delta function. Substituting \cref{eq:dirac} into \cref{eq:weakform} converts integrals to summation over material points. The coordinates of a material point located within an element is represented at time $t$ as

\begin{equation}
\mathbf{x}_p^t = \sum_{i=1}^{n_n} \mathbf{x}_i^t N_i(\mathbf{x}_p^t)
\label{eq:shapefunction}
\end{equation}

where spatial nodes are represented as $\mathbf{x}_i^t$ such that node number $i = (1,2...n_n)$ with $n_n$ representing the number of nodes in the element. The element type governs the applied basis function $N_i$. Additional variables also share this form of notation. For example, acceleration is given as:

\begin{equation}
\mathbf{a}_p^t = \sum_{i=1}^{n_n} \mathbf{a}_i^t N_i(\mathbf{x}_p^t)
\label{eq:acceleration}
\end{equation}

as is the weight function and velocity. Substituting \cref{eq:acceleration} and the summation form of the weak form \cref{eq:weakform}, the equation of motion in discretized form is written as

\begin{equation}
\label{eq:reduced1}
\sum_{j=1}^{n_n} m_{ij}^t\mathbf{a}_j^t = \mathbf{f}_i^{int,t} + \mathbf{f}_i^{ext,t}\,,
\end{equation}
\noindent where internal force vector is represented as

\begin{equation}
\mathbf{f}_i^{int,t} = -\sum_{p=1}^{n_p} m_p \sigma_p^{s,t} \cdot \nabla N_i(\mathbf{x})\vert_{\mathbf{x} = \mathbf{x}_p^t}\,,
\label{eq:internal}
\end{equation}

\noindent and the external forces is 

\begin{equation}
\mathbf{f}_i^{ext,t} = \sum_{p=1}^{n_p} m_p N_i(\mathbf{x}_p^t)\tau_p^{s,t} h^{-1} + \sum_{p=1}^{n_p}m_p\mathbf{b}_p^t N_i(\mathbf{x}_p^t) \,.
\label{eq:external}
\end{equation}

\noindent Which includes the addition of nodal mass matrix $m_{ij}^t$. The stress at a point is represented by

\begin{equation}
\label{eq:specstress}
\sigma_p^{s,t} = \sigma^s(\mathbf{x}_p^t,t)\,.
\end{equation}

The equation of momentum, \cref{eq:momentum} is a second-order differential equation with respect to time, so it can be solved by using either an explicit integration scheme or an implicit integration scheme. An implicit method involves finding the unknown displacement solution for the current time step by using information from both the current and previous time steps. An implicit time scheme is recommended for slowly applied and propagated stresses~\citep{kafaji13}. In contrast, an explicit time integration is commonly used to solve for the acceleration term. As the stresses experienced in a Slump-flow test are rapid, an explicit time scheme is chosen for this study. 

The stresses in the MPM algorithm can be updated before or after computing the acceleration term and are commonly referred to as separate methods termed updating stress first (USF) or updating stresses last (USL). \citet{bardenhagen02} observed that USL performed better than USF, in addition,~\citet{kafaji13} recommends the use of USL over USF in a granular flow-like simulation as USL dissipates energy in flow problems. However, the authors found that USF offered greater numerical stability with the Bingham constitutive model used in the study. All simulations discussed in this paper, therefore, use USF. In this study the authors developed a Thixotropic Bingham model which is implemented in the CB-GEO MPM code~\citep{soundararajan2019scalable}, described later in this section.

The choice of basis function, or shape function, $N_i$ can influence the accuracy and numerical stability of the MPM by reducing cell crossing noise. Cell crossing noise is caused by the discontinuity in point properties observed when a point crosses from one element to another. \citet{bardenhagen2004generalized} developed Generalised Interpolation Material Point (GIMP) to overcome cell crossing noise typically observed in linear MPM by using shape functions that span multiple elements. This GIMP method was implemented in the CB-GEO code based on~\citep{pruijn2016improvement} to minimise cell crossing errors in concrete flow problems. The GIMP was found to be especially effective in reducing the numerical error caused by points entering empty cells, thereby improving the stability of the simulation.

\subsubsection{Boundary Conditions}
A frictional boundary algorithm is used to describe the interaction between the concrete material points and the surface on which it flows. A description of this friction algorithm is available in~\citet{Samila}, which involves reducing the acceleration of a point based on a Friction Coefficient (FC). The FC in the Slump-Flow test represents the kinetic and static friction between the concrete and the steel plate. It is well documented in the literature that FC is influenced by rheology, mix design (such as cement/water ratio) and surface roughness of the plate. \citet{sofcf} describes friction as a key variable for simulating concrete flow. However, a wide variation in the suitable values of FC can be observed in the literature. For simulating concrete with an SF of \SI{800}{\mm},~\citet{Deeb2014} recommended a dynamic FC of $0.55$. While~\citet{djelal2004tribological} reported a FC of $0.06$-$0.1$ for concrete with an SF of \SI{700}{\mm}. For an L-box test, which is similar to the SF test,~\citet{HOSSEI17} observed good correlation with experimental results for an FC of $0.4$. As in these cases, FC is often used as a calibration parameter in simulations of concrete flow. 

\citet{pviscosity_friction} observed that the volume of cement paste and the amount of superplasticiser within a concrete mix affects the friction at the concrete plate interface. The friction stresses at the interface were observed to increase with increasing paste volume due to the enrichment of the boundary layer caused by the availability of more fines. The greater thickness of the boundary allows more cement grains to be trapped in the roughness of the metal plate. \citet{pviscosity_friction} also demonstrated that friction stress, which is proportional to the friction coefficient, decreases with increasing plastic viscosity. In the present study, calibrated FC values between $0.35$ and $0.45$ are used to represent the friction between Tremie Concrete and the base-plate, as shown in \cref{table:table1}. FC also increases with decreasing plastic viscosity to align with the observations made in \citet{pviscosity_friction}.

\subsubsection{Geometry} 
The geometry of the Sump-Flow test in 3D is discretised using approximately $17,000$ material points. $4000$ points represent the Slump-cone with the remainder representing Tremie Concrete. High density, rigid material points are used to model the Slump-cone to prevent any unwanted outward flow of concrete from the cone. It is important to simulate the cone as it confines the initial flow of concrete, reducing the overall spread. Without the cone, the simulations are not subject to the correct confining forces, producing an unrealistic early geometry of the flow. At the start of the simulation, the cone is raised at a speed of \SI{15}{\cm\per\second} so as to raise the cone in $1$-\SI{3}{\second} as per the~\citet{BS123508} guidelines. The background mesh is discretised with $256,000$ nodes creating $219,373$ hexahedral GIMP elements.

\subsection{Constitutive Model}
According to \citet{reddy2007}, in 3 dimensions, the gradient of the velocity $u$ is a second-order tensor which can be expressed as the matrix $\mathbf{L}_{ij}$:

\begin{equation}
\mathbf{L}_{ij}=\frac{\partial u_i}{\partial x_j}.
\label{eq:L}
\end{equation}

\noindent $\mathbf{L}_{ij}$ can be decomposed into the sum of a symmetric matrix 
$\mathbf{D}_{ij}$ and a skew-symmetric matrix $\mathbf{W}_{ij}$ as follows

\begin{equation}
\mathbf{D}_{ij} = \frac{1}{2} (\mathbf{L}_{ij} + \mathbf{L}_{ij}^T) 
\label{eq:d1}
\end{equation}
\begin{equation}
\mathbf{W}_{ij} = \frac{1}{2} (\mathbf{L}_{ij} - \mathbf{L}_{ij}^T)
\label{eq:w1}
\end{equation}

\noindent where $\mathbf{D}_{ij}$ and $\mathbf{W}_{ij}$ are the rate of deformation tensor:

\begin{equation}
 \mathbf{D}_{ij}=\left[ \begin{matrix} \frac{du_x}{dx} & \frac{1}{2}(\frac{du_x}{dy} + \frac{du_y}{dx}) & \frac{1}{2}(\frac{du_x}{dz} + \frac{du_z}{dx}) \\ \frac{1}{2}(\frac{du_x}{dy} + \frac{du_y}{dx}) & \frac{du_y}{dy} & \frac{1}{2}(\frac{du_y}{dz} + \frac{du_z}{dy}) \\ \frac{1}{2}(\frac{du_x}{dz} + \frac{du_z}{dx}) & \frac{1}{2}(\frac{du_y}{dz} + \frac{du_z}{dy}) & \frac{du_z}{dz}
\end{matrix} \right] 
\label{eq:dtensor}
\end{equation}

\noindent and the vorticity tensor or spin tensor respectively. In the CB-GEO MPM, the strain rate tensor $\dot\epsilon$ is calculated as $\mathbf{B} u$, where $\mathbf{B}$ is the strain-displacement B-matrix \citep{zienkiewicz2005} to give

\begin{equation}
 \dot \epsilon= \mathbf{B}u =\left[ \begin{matrix} \frac{du_x}{dx} & (\frac{du_x}{dy} + \frac{du_y}{dx}) & (\frac{du_x}{dz} + \frac{du_z}{dx}) \\ (\frac{du_x}{dy} + \frac{du_y}{dx}) & \frac{du_y}{dy} & (\frac{du_y}{dz} + \frac{du_z}{dy}) \\ (\frac{du_x}{dz} + \frac{du_z}{dx}) & (\frac{du_y}{dz} + \frac{du_z}{dy}) & \frac{du_z}{dz}
\end{matrix} \right]
\label{eq:epsilondot}
\end{equation}

\noindent such that the rate of deformation tensor can also be represented as a $6$x$1$ function of the rate of strain 

\begin{equation}
\mathbf{D}_{ij} = \left[ \dot \epsilon_{xx}, \dot \epsilon_{yy}, \dot \epsilon_{yy}, \frac{1}{2}\dot \epsilon_{xy}, \frac{1}{2} \dot \epsilon_{yz}, \frac{1}{2} \dot \epsilon_{zx}\right]^T \,.
\label{eq:dstrain}
\end{equation}

\noindent Using the rate of deformation, \cref{eq:bingham} can be expressed as

\begin{equation}
\boldsymbol\tau_{ij} = 2 \left ( \frac{\tau_0}{\dot{\gamma}} + \mu \right )\mathbf{D}_{ij} \quad \textrm{if} \quad |\boldsymbol\tau_{ij}|\geq \tau_0
\label{eq:bingham_2}
\end{equation}

\noindent where $\boldsymbol\tau_{ij}$ is the deviatoric stress tensor and $\mathbf{D}_{ij} = 0$ if $|\boldsymbol\tau_{ij}| < \tau_0$. The shear rate $\dot{\gamma}$ can now be described as

\begin{equation}
\dot{\gamma} = (2\mathbf{D}_{ij}\mathbf{D}_{ij})^{1/2}\,.
\label{eq:shearate}
\end{equation}

The total stress $\sigma_{ij}$ can be calculated from the deviatoric stress tensor, the thermodynamic pressure $P$ and the Dirac Delta function $\delta_{ij}$ by:

\begin{equation}
\boldsymbol\sigma_{ij} = -P\boldsymbol\delta_{ij} + \boldsymbol\tau_{ij}
\label{eq:total_stres}
\end{equation}

\noindent where pressure $P$ is 

\begin{equation}
P = p_0 - Kd\epsilon_v\,.
\label{eq:pressue}
\end{equation}

\noindent Here, $p_0$ is the initial thermodynamic pressure (calculated as the mean initial stress) $d\epsilon_v$ is the volumetric strain and $K$ is the bulk modulus 
\begin{equation}
K = \frac{E}{3(1-2v)}
\label{eq:bulk}
\end{equation}
In this study, we use a Young's Modulus $E$ of \SI{0.01}{\MPa}~\citep{roussel2012origins} and a Poisson's ratio $v$ of $0.45$ to represent a weakly incompressible material. The thermodynamic pressure is calculated at the centre of the MPM cell to minimise pressure oscillations. 

Fresh concrete modelled using a viscoplastic Bingham material behaves like a rigid body before yielding, and as a viscous material upon yielding. In order for the material to transition from rigid to viscous, the magnitude of the deviatoric stress tensor must exceed the yield stress. Based on \cref{eq:bingham_2}, total stresses will remain zero when the material is in the rigid region because the rate of deformation will also be zero. In reality, this is not physically possible as a material in the rigid zone will still exhibit non-zero stresses, despite not flowing. Hence, a modification to this model is needed to allow non-zero stresses to develop before the material yields.

One such modification, or regularisation, is to define the behaviour before and after yield as two separate materials. The most commonly adopted modification is the Bi-viscous model~\citep{BEVERLY199285}, which uses a critical shear rate $\dot{\gamma}_c$ to define the shear rate at which the transition between rigid and viscous behaviour occurs. The Bi-viscous model is written as:

\begin{equation}
\boldsymbol\tau_{ij}= 
\begin{cases}
  2\mu_0 \mathbf{D}_{ij} & \text{if } |\dot{\gamma}| \leq \dot{\gamma}_c\\
   2 \left ( \frac{\tau_0}{\dot{\gamma}} + \mu \right )\mathbf{D}_{ij} & \text{otherwise}
\end{cases}
\label{eq:bivis}
\end{equation}

where $\mu_0$ represents the viscosity of the material in the pre-yield region, and typically has a value three orders of magnitude higher than the dynamic viscosity $\mu$.

The Bi-viscous model offers a clear differentiation between the yielding and pre-yield regions by using a critical yield stress. However, the abrupt discontinuity at the transition between the regions still presents a computational hurdle. An alternative regularising method involves adopting a continuous model which does not have a distinct discontinuity between rigid and plastic regions. The Papanastasiou~\citep{pap87} model represents the most popular of these kinds of continuous models. The exponential regularisation of the Bingham model often referred to as the Bingham-Papanastasiou model, introduces a parameter $m$ which controls the exponential growth of stress. The 3D representation of this regularisation is:

\begin{equation}
\boldsymbol\tau_{ij}= 2\left[ \mu + \frac{\tau_0}{\dot{\gamma}}(1-e^{-m \dot{\gamma}})
\right]\mathbf{D}_{ij}
\label{eq:pap1}
\end{equation}

The larger the regularisation parameter $m$, the closer the numerical representation will be to the Bingham model in \cref{eq:bingham_2}. However, any increase in $m$ will also result in a significant increase in the apparent viscosity at low strain rates. Thus, high $m$ values create an unrealistic stiffness which can lead to volumetric locking within MPM. Locking effects in MPM can be mitigated by reducing the number of material points in cells and calculating the volumetric deformation at the centre of the cell~\citep{coombs2018overcoming}. Due to the significant increases in stiffness exhibited by thixotropic concrete and the use of velocity gradient rather than deformation gradient in the MPM implementation, these mitigations were not sufficient to reduce the effect of locking to a suitable level. \citet{franci20183d} demonstrated how different $m$ values have a marginal impact on the behaviour of a Bingham representation of cement paste but can influence the stability of a model. An $m$ value of 5 reduced the influence of locking on the simulations substantially and is subsequently used in all simulations in this study. For the first time, we propose a new model that combines Roussel's thixotropic equations which describe time-dependent concrete behaviour with a continuous Bingham model by substituting \cref{eq:thix_int} into \cref{eq:pap1} to give:

\begin{equation}
\boldsymbol\tau_{ij}= 2\left[ \mu + \frac{(1+\lambda_0e^{-\alpha\dot{\gamma}t})\tau_0}{\dot{\gamma}}(1-e^{-m \dot{\gamma}})
\right]\mathbf{D}_{ij} \,.
\label{eq:pap2}
\end{equation}

This model will be referred to as the the Papanastasiou-Roussel Bingham (PR-Bingham) model for the remainder of this paper. \Cref{eq:pap2} is valid for all strain-rates and degrees of concrete thixotropy. However, for computational purposes, there is still a need to define a point at which thixotropy increases whilst the material is at rest, and a point where thixotropy begins to decrease as the material flows. In this work, a critical shear rate, $\dot{\gamma}_c$, is used to define a yield point, similar to the approach taken by~\citet{BEVERLY199285}. In the model adopted here, any material with a shear rate below the critical shear rate is considered to be at rest, allowing thixotropy to build. However, material at these low shear rates may still exhibit a degree of movement as a result of using the Papanastasiou model. From an engineering perspective the material can still be considered at rest even with a small, often negligible, degree of deformation as noted by~\citet{franci20183d}. 

\citet{roussel16} suggested the use of a stopping criterion which allows for the definition of the final shape of the free surface of the material regardless of any continued, negligible deformation. This criterion has taken the form of a flow-cut off time in previous studies~\citep{claud_thesis}, where it is assumed that any flow occurring after this point is a result of the limitations of the constitutive model. The stopping criterion used in this study is \SI{20}{\second} of flow time, in line with that used by~\citet{claud_thesis}. This represents more than the maximum SF time for all concretes presented in \cref{fig:thix_comp}~\citep{krankel,dfi_feys}.

\begin{figure}
\centering
\includegraphics[width=0.8\linewidth]{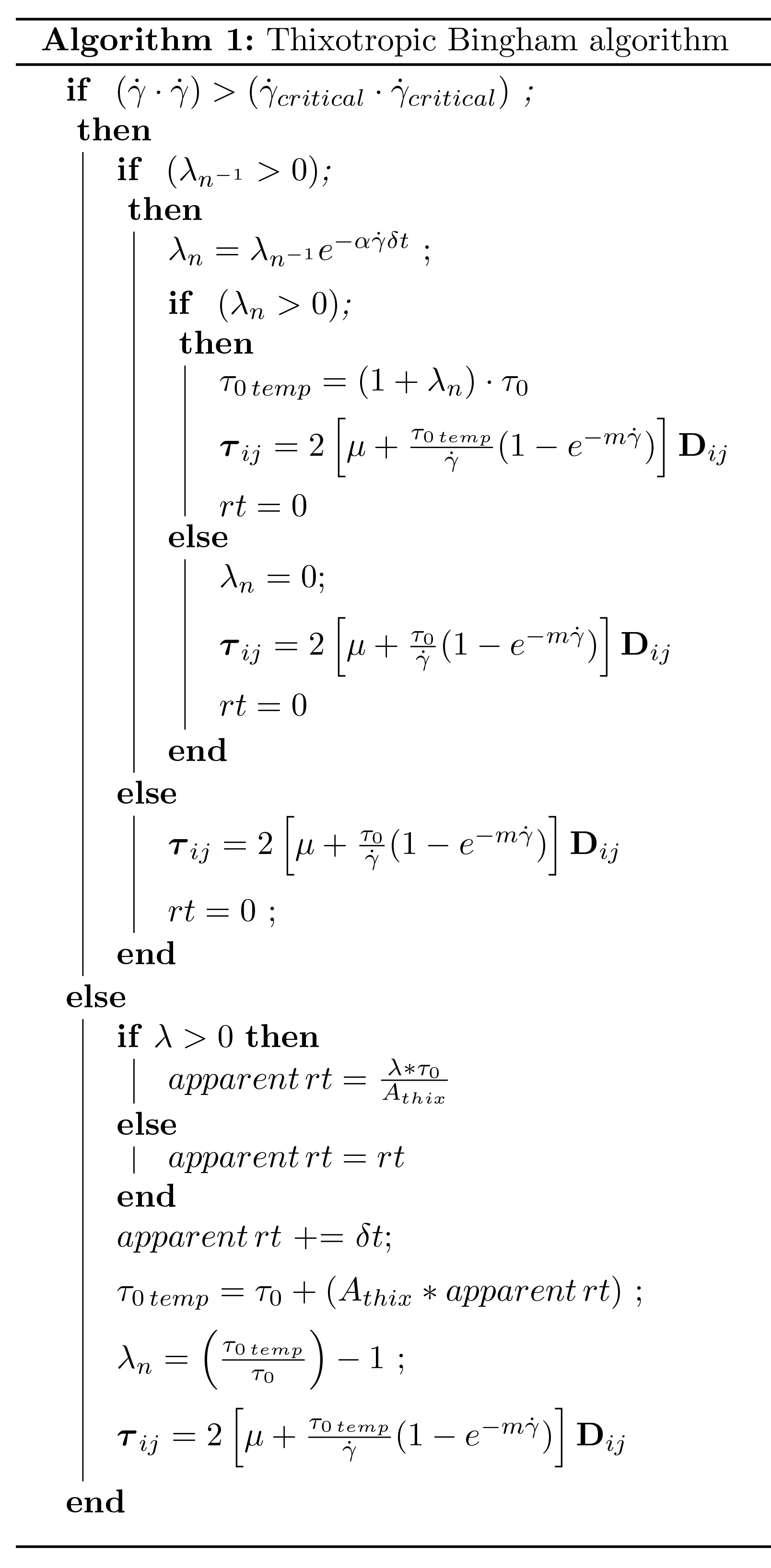}
\caption{\label{fig:algo1} The Papanastasiou-Roussel Bingham (PR-Bingham) algorithm implemented in CB-GEO MPM.}
\end{figure}

~\Cref{fig:algo1} shows the Algorithm used to build up thixotropic stresses in CB-GEO MPM in accordance with~\cref{eq:pap2}, where $\delta t$ represents the length of a time-step (\SI{1.0E-4}{\second}), $\lambda_{n^{-1}}$ the flocculation state of the previous time-step, and the addition of a yield criterion in the form of a critical shear rate. All results in the following section use this algorithm to calculate the stress state of the material over time.

\begin{figure}
\centering
\includegraphics[width=\linewidth]{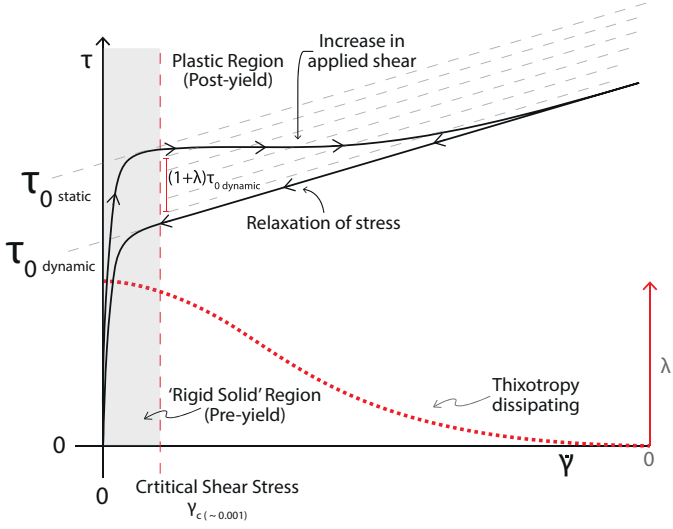}
\caption{\label{fig:thix_graph}1D Bi-viscous Thixotropic Bingham hysteresis loop}
\end{figure}
An appropriate way to characterise a thixotropic system is to use the hysteresis loop of the flow curve, representing an applied shear stress (upward) followed by a relaxing of the stress (downward)~\citep{Baltazar19}.
\Cref{fig:thix_graph} represents a 1D single hysteresis loop analogous to~\cref{fig:algo1}, demonstrating how the thixotropy multiplier $\lambda$ increases the yield stress after a period of rest and dissipates to $0$ during shearing, resulting in a hysteresis loop.

\begin{figure}
\centering
\includegraphics[width=\linewidth]{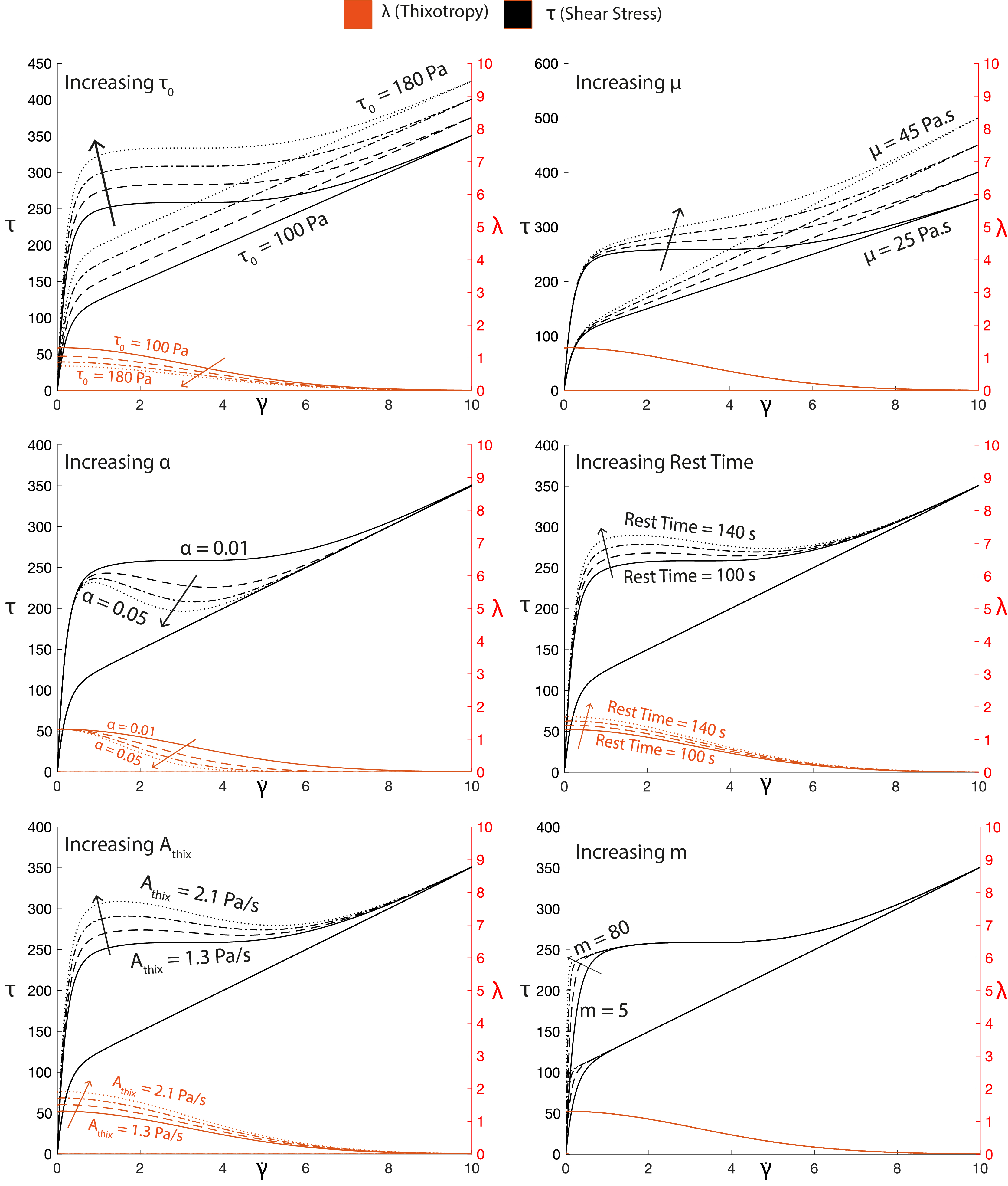}
\caption{\label{fig:parametric}Parametric study of thixotropic algorithm. Blue represents the shear stress and orange the thixotropic multiplier $\lambda$.}
\end{figure}

A parametric study is performed to understand the effect of the model parameters ($\tau_0$, $\mu$, $\alpha$, $A_{thix}$, $m$ and rest time) on the shear stress and the thixotropic multiplier $\lambda$ following a \SI{100}{\second} rest period. The results of the parametric study are presented in~\cref{fig:parametric}. An increase in $\tau_0$ causes an upward translation of the shear stress intercept, and a downward translation of $\lambda$ as the constant increase in yield stress over time is now proportionately smaller than the initial yield stress. Increasing $\mu$ sees the gradient of the $\tau$/$\dot{\gamma}$ curve steepen, creating higher stresses at lower shear rates. Changes in rest time and $A_{thix}$ both increase the $\tau_{0_{static}}$ reached after a period of rest. A larger $\alpha$ value causes a reduction in the time required for the elevated $\tau_{0_{static}}$ to reduce to the $\tau_{0_{dynamic}}$ during shearing. Finally, as discussed previously, a larger $m$ results in a model closer to the true Bingham model, but at a cost of increasing the stresses at lower shear rates.

\subsection{Calibration of input parameters}

\begin{table}
\centering
\caption{\label{table:table1} MPM input parameters of Mixes A-C, based on~\citep{krankel,dfi_feys}}
\begin{tabular}{lcccccc}
 %     & \multicolumn{1}{l}{$\tau_{0_{dynamic}}[Pa]$} & \multicolumn{1}{l}{$\mu [Pa\cdot s]$} & 
 %     \multicolumn{1}{l}{$A_{thix}[Pa s^{-1}]$} & \multicolumn{1}{l}{SF$_{0}\pm30$[mm]} & 
 %     \multicolumn{1}{l}{SF$_{240}\pm 30$[mm]} & \multicolumn{1}{l}{FC} \\ 
  & \multicolumn{1}{l}{$\tau_{0_{dynamic}}$} & \multicolumn{1}{l}{$\mu $} & 
      \multicolumn{1}{l}{$A_{thix}$} & \multicolumn{1}{l}{SF$_{0}\pm30$} & \multicolumn{1}{l}{SF$_{240}\pm 30$} & \multicolumn{1}{l}{FC} \\ 
            & \multicolumn{1}{c}{$[Pa]$} & \multicolumn{1}{c}{$[Pa\cdot s]$} & 
      \multicolumn{1}{c}{$[Pa s^{-1}]$} & \multicolumn{1}{c}{[mm]} & \multicolumn{1}{c}{[mm]} &  \multicolumn{1}{l}{} \\ 
\hline
Mix A & 54           & 35.8         & 1.3            & 590            & 575       &   0.35    \\
Mix B & 61           & 29          & 1.8             & 550            & 530      &   0.45    \\
Mix C & 130           & 30.2         & 2.9            & 500            & 485       &   0.43  
\end{tabular}
\end{table}

To demonstrate the capability of the developed PR-Bingham MPM model in simulating Tremie Concrete, the results of numerical simulations are compared against experimental slump-flow tests~\citep{krankel,dfi_feys}. Dynamic yield stress, plastic viscosity, $A_{thix}$, $SF_{0}$ and SF$_{240}$ values were obtained from an EFFC and DFI data-set~\citep{krankel,dfi_feys} of concrete tests performed on Tremie Concrete from numerous European and North American sites. Three concretes from this data set, referred to as Mix A-C, are chosen based on the range of concrete properties they represent. The model parameters for Mixes A-C used in this study are shown in \cref{table:table1}, where SF$_{0}$ and SF$_{240}$ represent the measured SF in \si{mm} for $0$ seconds of rest and $240$ seconds of rest, respectively, for the physical test.

Mix A is a low yield stress, high viscosity concrete with a low level of thixotropy. This produces a SF of almost \SI{600}{\mm} which is in the upper range of acceptability. Mix B has a similar yield stress to Mix A, but has a significantly lower viscosity, resulting in lower FC requirement and a mid-range SF of \SI{550}{\mm}. Mix B also has a slightly higher level of thixotropy, as represented by the increase in $A_{thix}$ of 0.5 from Mix A. Finally, Mix C has a significantly higher yield stress, resulting in a SF of only \SI{500}{\mm}, near the lower limit of acceptability. Mix C also has a significantly higher degree of thixotropy, with an $A_{thix}$ of $2.9$. The SF behaviour of the three different mixes and their comparison with the physical experiments are presented in the following section.

Furthermore, we evaluate the influence of incorporating thixotropy to accurately represent concrete. A rest time of \SI{240}{\second} is introduced in the PR-Bingham algorithm to induce time-dependent changes in the concrete as described in~\cref{fig:algo1}. Depending on the duration of the rest time, an appropriate increase in the $\lambda$ value causes an increase in the shear stress to $\tau_{0_{static}}$.

\section{Results and Discussion} 
\label{section:results}
\subsection{Dynamic Property Simulations}
\begin{figure}
\centering
\includegraphics[width=\linewidth]{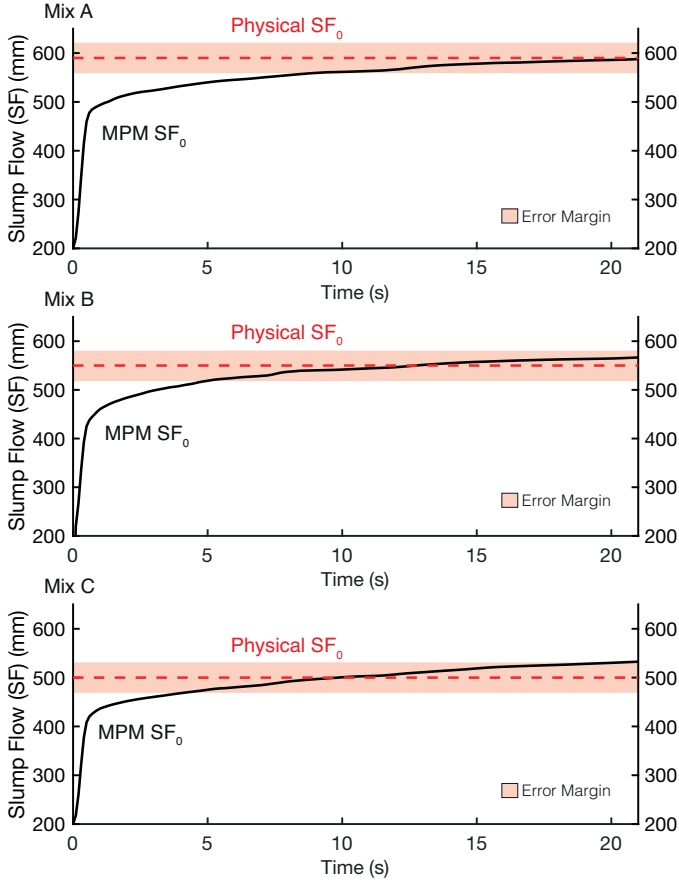}
\caption{\label{fig:sf0} Evolution of MPM simulated SF$_{0}$ over time for each concrete mix, dashed line indicates the physical SF$_{0}$ and MPM simulated SF$_{0}$ from \cref{table:table1}.}
\end{figure}

\Cref{fig:sf0} shows the time-evolution of the Slump-flow spread in \si{\mm} for Mixes A-C (see~\cref{table:table1}) where no rest time is present. For Mix A, a highly viscous low yield stress concrete, the MPM $SF_{0}$ reaches the corresponding Physical SF$_{0}$ around \SI{18}{\second}. This is expected for a high viscosity concrete like Mix A. The MPM simulations of Mixes B and C reach the physical SF at approximately \SI{10}{\second} and \SI{12}{\second} respectively, as both mixes have a lower viscosity in comparison to Mix A. The MPM SF$_{0}$ for Mixes A and B continues to grow at a negligible level after reaching the physical SF$_{0}$ due to low shear deformation caused by PR-Bingham model. Despite this continued growth, the MPM SF$_{0}$ for mixes A and B remain within the margin of error~\citep{long_rep}. The MPM SF$_{0}$ for Mix C does continue to grow at a higher rate after reaching the physical SF$_{0}$, but still reached a stable level within the physical SF$_{0}$ margin of error. The continued growth of Mix C is due to the constant deformation at a low shear rate characteristic of the PR-Bingham model with larger yield stress.

\begin{figure*}
\centering
\includegraphics{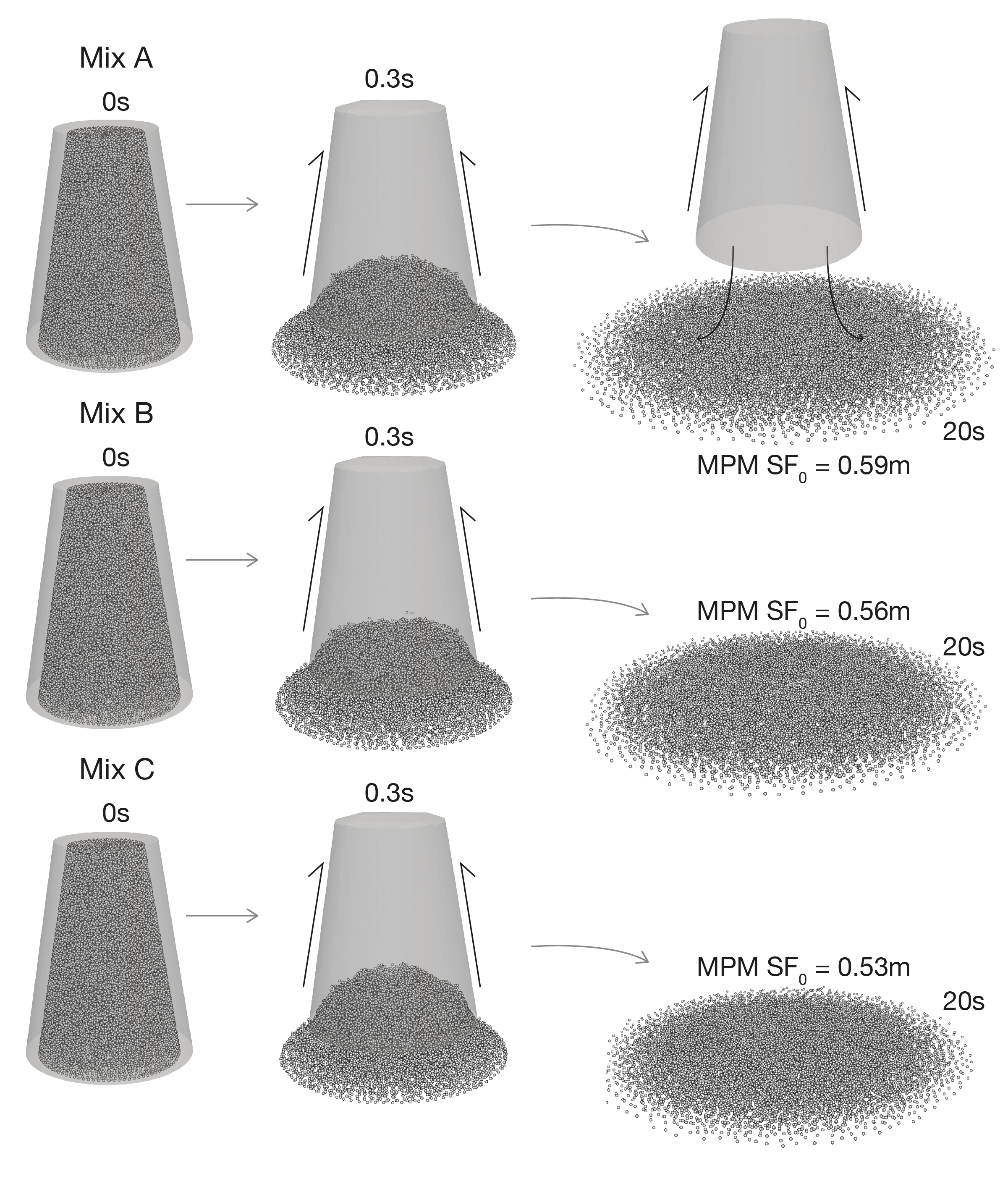}
\caption{\label{fig:mpm_54}Time evolution of Slump-flow test for all mixes. From top to bottom: Mix A, Mix B and Mix C}
\end{figure*}

\Cref{fig:mpm_54} shows the three-dimensional isometric view of the evolution  of all three mixes as they leave the cone at \SI{0.3}{\second} and the final shape at \SI{20}{\second}. \Cref{fig:mpm_54} shows a clear difference in spread diameter for the mixes at \SI{0.3}{\second}. The final spread are significantly different, but the three mixes maintain a uniform shape and thickness.

The good level of correlation between the physical and numerical SF$_{0}$ indicates that the PR-Bingham model developed in the MPM framework is capable of modelling the rheological behaviour of Tremie Concrete. 
\subsection{Thixotropic Simulations}

\begin{figure}
\centering
\includegraphics[width=\linewidth]{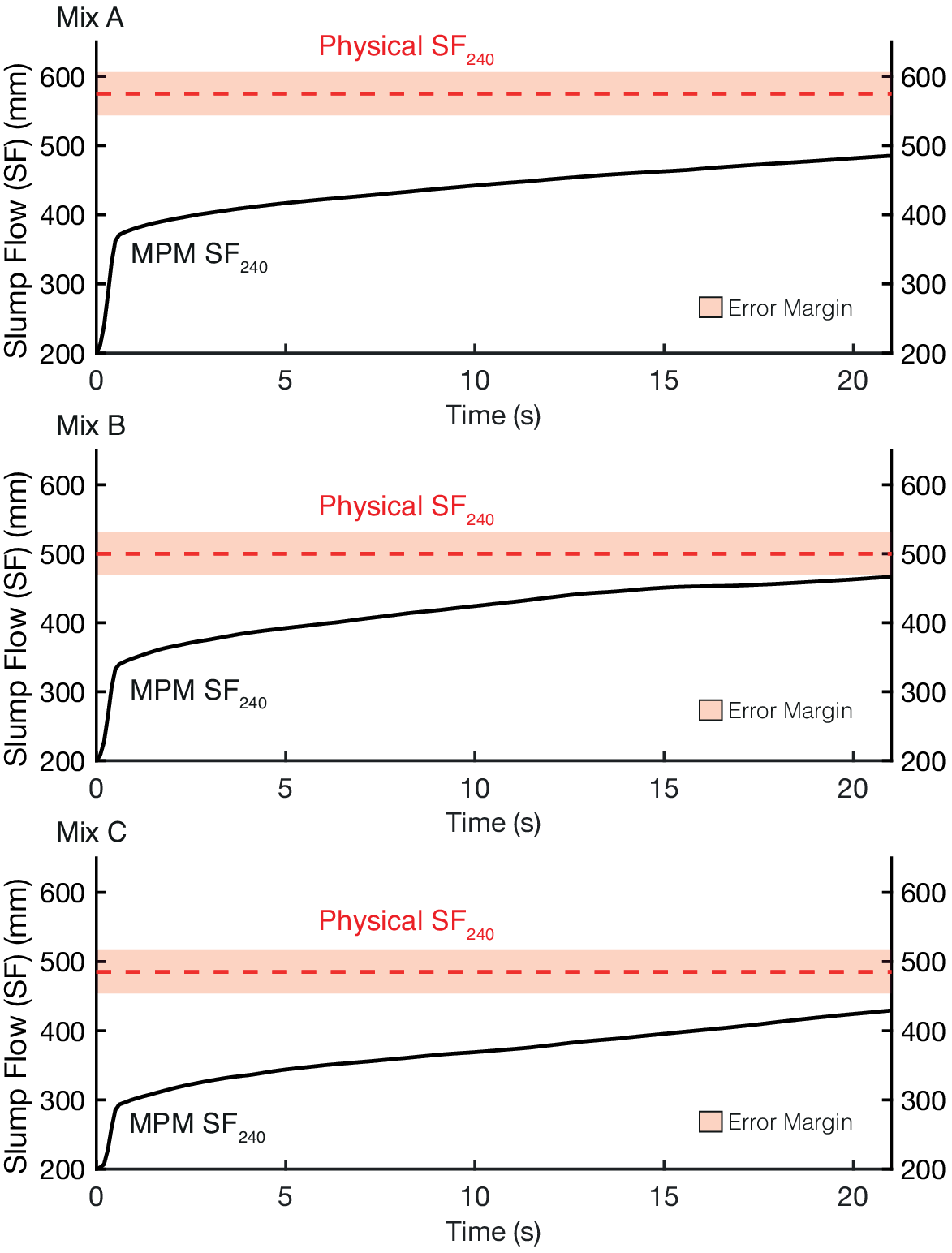}
\caption{\label{fig:sf240} Evolution of MPM simulated SF$_{240}$ over time for each concrete mix, dashed line indicates the physical SF$_{240}$ from \cref{table:table1}.}
\end{figure}

The time dependent thixotropic behaviour of Tremie concrete is simulated by introducing a rest time to build the yield stress of Tremie Concrete.~\Cref{fig:sf240} compares the evolution of the MPM SF$_{240}$ and the Physical SF$_{240}$ for Mixes A-C. After an initial rest period of \SI{240}{\second} all three concretes fall short of reaching the final experimental SF by approximately $60$-\SI{80}{\mm}.

\begin{table}
\centering
\caption{\label{table:dsf} Difference in Slump-flow following a period of rest for physical tests and MPM simulations.}
\begin{tabular}{lccll}
\toprule
      & \multicolumn{2}{c}{$\Delta$SF (mm)} & \multicolumn{2}{l}{$\Delta$SF ($\%$ of SF$_0$)} \\ \cmidrule{2-5}
      & Physical            & MPM           & Physical               & MPM               \\ \midrule
Mix A & 15                  & 105           & 2.6                      & 18.0                 \\
Mix B & 20                  & 94            & 4.0                      & 18.8                 \\
Mix C & 15                  & 98            & 3.1                      & 20.2                 \\ \bottomrule
\end{tabular}

\end{table}

The difference in the SF run-out with and without the initial rest time of \SI{240}{\second} is defined as $\Delta$SF. The $\Delta$SF values for the MPM simulations are summarised in \cref{table:dsf} for Mixes A-C, showing an approximately \SI{100}{\mm} change in the SF following a \SI{240}{\second} rest. Conversely, as discussed earlier in this paper, the experimental $\Delta$SF values for Mixes A-C measured in the physical tests are significantly lower, ranging between \SI{15}{\mm} and \SI{20}{\mm}. In the PR-Bingham model, the values of $A_{thix}$ over $1$ cause an increase in the yield stress of over \SI{240}{\pascal} during a rest period of \SI{240}{\second}. \Cref{fig:t0_sf}  shows that an increase in yield stress of over \SI{240}{\pascal} should yield a significant $\Delta$SF, more than the observed $15$-\SI{20}{\mm} difference. The $\Delta$SF observed in MPM simulations are still lower than what is the expected given the generalised empirical relationship between SF and $\tau_0$ for a concrete with a yield stress of $>$\SI{250}{\pascal}, but is closer to the expected behaviour than the $\Delta$SF Physical values. 

As discussed in \cref{section:thix},~\citet{krankel, dfi_feys} could not find a consensus on why the physical Slump-flow test could not accurately represent thixotropic concrete, just that it was inadequate at doing so. This is supported by the small $\Delta$SF physical values in \cref{table:dsf}. Interestingly, the MPM simulations appear to capture a behaviour closer to the expected flow behaviour, with large $\Delta$SF values reported in the MPM simulations. It is possible that the simulations overcome some physical artefacts that prevents the physical Slump-Flow test from representing the difference in the flow behaviour of thixotropic concrete. The MPM simulations also align better with the rheometer measurements of Mixes A-C reported by~\citep{krankel, dfi_feys}, which describes large increases in yield stress (material stiffness) over short rest periods, as noted by the high $A_{thix}$ values in \cref{table:table1}.

\begin{figure}
\centering
\includegraphics[width=\linewidth]{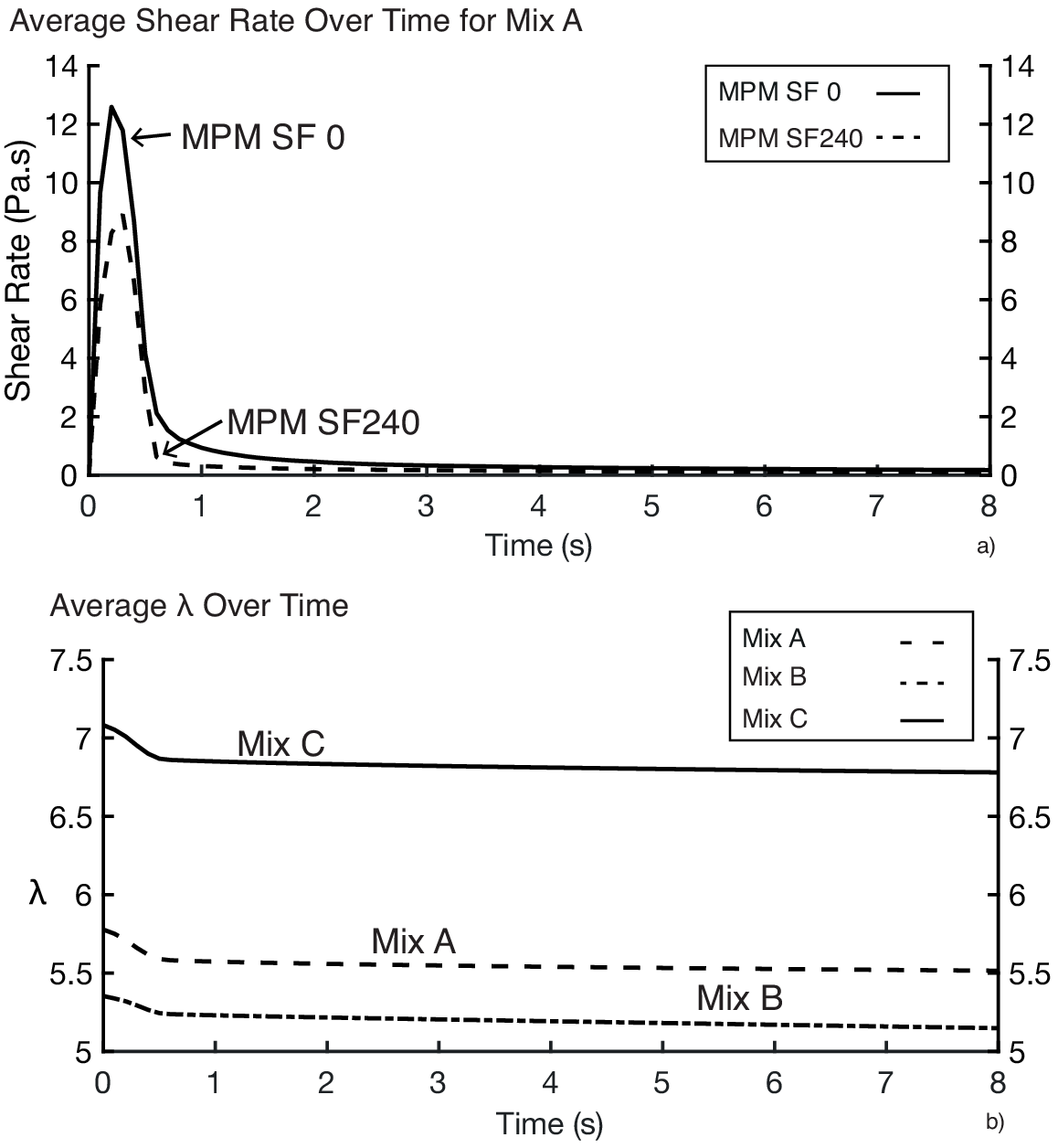}
\caption{\label{fig:lambda_shear} A) Mean average shear rate for Mix A during the Slump-Flow test for \SI{0}{\second} rest and \SI{240}{\second} rest  B)Time evolution of $\lambda$ for all mixes following a rest period of \SI{240}{\second}. From top to bottom: Mix B, Mix A and Mix C}
\end{figure}

It is possible that high shear rates during the slump-flow test remove a significant portion of thixotropy, preventing the test from accurately representing thixotropic concrete. This proposition can be disproved by examining the mean average shear rate occurring during a Slump-flow test, \cref{fig:lambda_shear}a. In Mix A, the peak shear rate is approximately \SI{12}{\per\second} and \SI{9}{\per\second} for \SI{0}{\second} rest and \SI{240}{\second} rest, respectively. However, the slump-flow test experiences this peak shear rate only for a very short duration and is not sufficient to dissipate the large amount thixotropy developed at rest. The shear-stress that develops as the rate of SF spreading declines will also not be sufficient to remove a significant portion of thixotropy. This can be further supported by the time required to reduce static shear stresses to dynamic in a rheometer, which is around \SI{10}{\second} and requires a constant shear-rate of up to \SI{12}{\per\second}~\citep{ICAR}. The Slump-flow test does not experience shear rates high enough to dissipate much thixotropy. The implication of this observation is that if the slump flow test does not experience enough shear stresses to break down thixotropy, then the $\Delta$SF for the physical test should be much larger, as it is with the MPM $\Delta$SF. This supports the observation that MPM is not only capable of simulating dynamic concrete properties, but also static properties of thixotropic concrete.

\begin{figure*}
\centering
\includegraphics{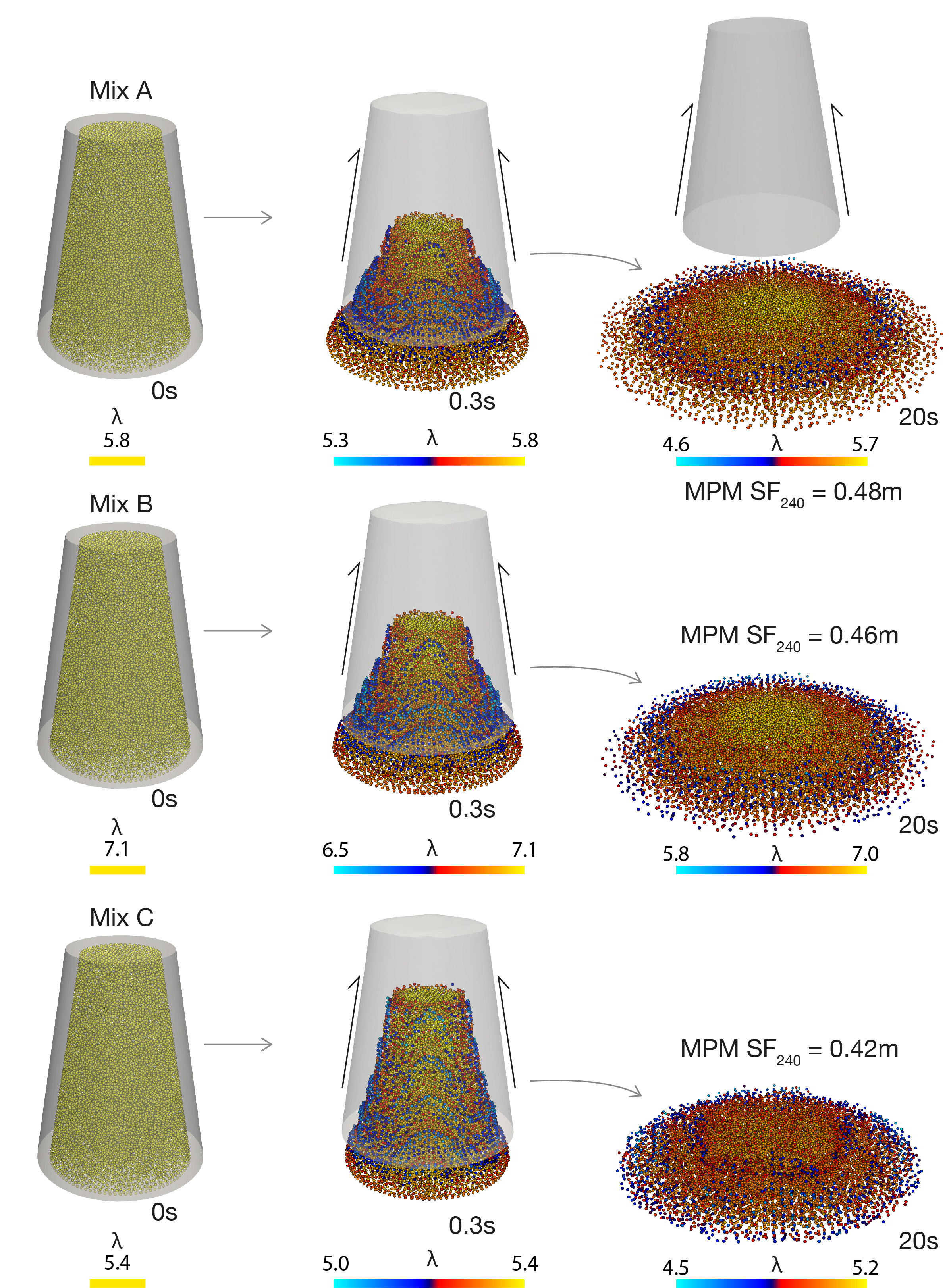}
\caption{\label{fig:lambda_all} Time evolution of Slump-flow test for all mixes following a rest period of \SI{240}{\second}. From top to bottom: Mix A, Mix B and Mix C}
\end{figure*}

The thixotropic parameter $\lambda$ directly controls the increase in the yield stress.~\Cref{fig:lambda_all} shows the dissipation of $\lambda$ with time for all mixes. At \SI{20}{\second} of flow-time, the result of the relatively low shear rates experienced by the concrete has prevented any notable dissipation of thixotropy. As the lowest shear rates are experienced close to the centre of the spread, the thixotropic dissipation has occurred the least in these areas. For all three mixes shown in \cref{fig:lambda_all}, the centre regions display  higher $\lambda$ values than the edges. This again supports the argument that the Slump-flow test is not capable of dissipating a significant amount of thixotropy.

\Cref{fig:lambda_shear}b reveals the rate of dissipation of the mean $\lambda$ over time during the test for all three mixes. It can be noted here that from approximately \SI{3}{\second} of flow time, the gradient of the slope remains relatively constant. Therefore, the bulk of thixotropy dissipation is likely to happen within the initial period of the test.

\begin{figure}
\centering
\includegraphics[width=\linewidth]{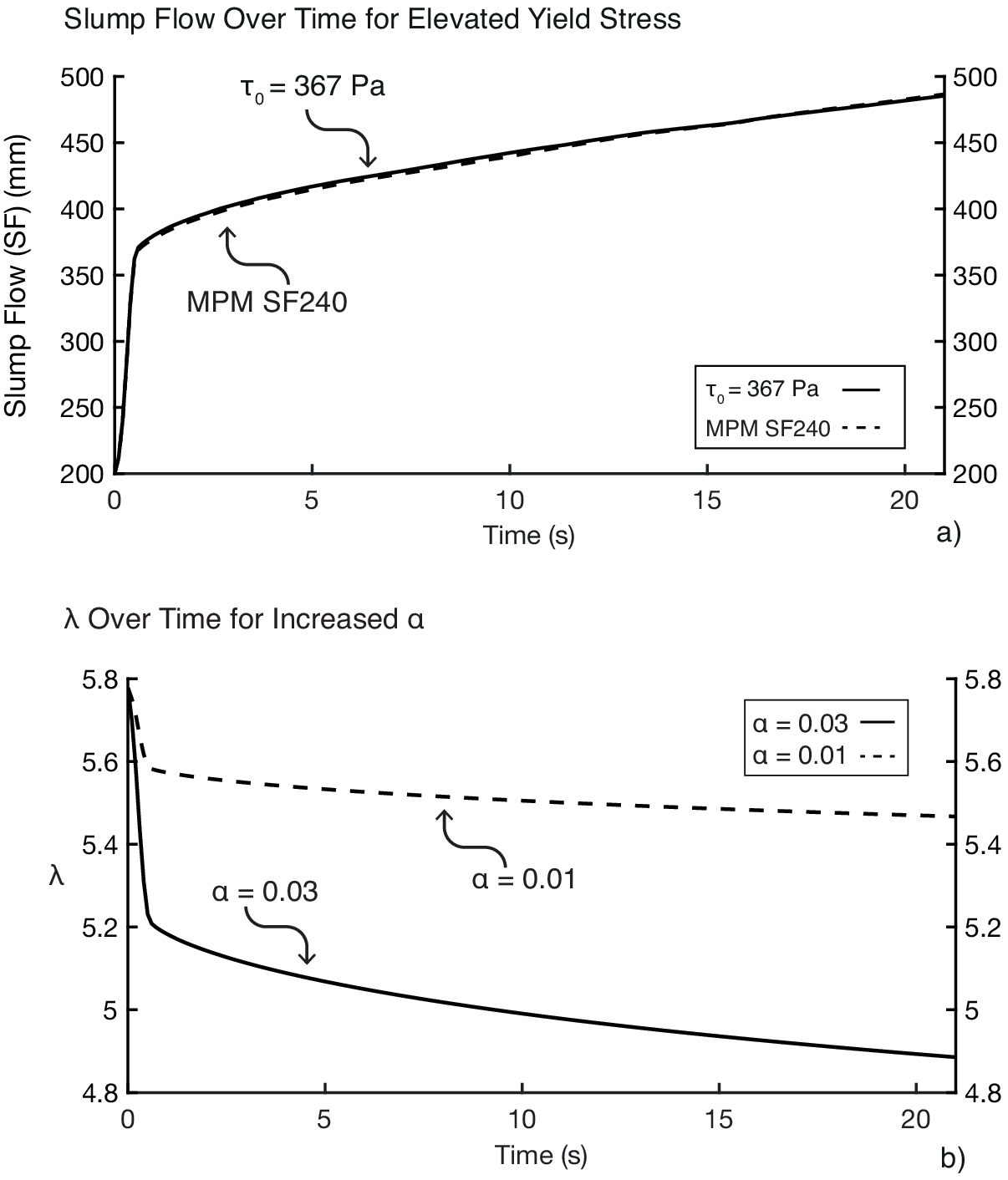}
\caption{\label{fig:367}A) Time evolution of Slump-flow test for Mix A following a rest period of \SI{240}{\second} and from \SI{0}{\second} rest but with a starting yield stress equal to Mix A at \SI{240}{\second} rest. B) Effect of alpha increase on dissipation of $\lambda$.}
\end{figure}

One final piece of evidence suggesting the inability of the Slump-flow test to dissipate thixotropy is presented by \cref{fig:367}a. Here, a simulation is conducted where all starting properties are equal to Mix A MPM SF$_0$, apart from the starting yield stress, which has been increased to \SI{367}{\pascal}. This value is equal to the static yield stress of Mix A following a \SI{240}{\second} rest. The final SF measurements and the evolution of SF are almost identical, as shown by the overlain curve (dashed line) of Mix A MPM SF$_{240}$. This suggests that the thixotropic effects developed at rest within concrete  are not dissipating, resulting in an elevated yield stress that affects the final run-out distance.

In the previous section of this paper, a parametric study (\cref{fig:parametric}) of the PR-Bingham parameters was shown. The parameter, $\alpha$, influenced the rate at which thixotropy would dissipate following the application of stress. For the MPM simulations presented, an $\alpha$ of $0.01$ was used, to be inline with experimental observations from a rheometer and observations in the literature. To provide continued evidence that the Slump-flow test does not dissipate thixotropy significantly, \cref{fig:367}b demonstrates the effect on thixotropy dissipation for two different values of alpha ($\alpha = 0.03$ and $0.01$). What can be observed, for Mix A, is that an increase in alpha from $0.01$ to $0.03$ causes a reduction of thixotropic multiplier $\lambda$ from $5.7$ to approximately $4.8$, as opposed to approximately $5.4$ for an $\alpha$ of $0.01$. While this does represent an increase in thixotropy dissipation, the final SF was within \SI{5}{\mm} of Mix A MPM SF$_{240}$. Thus, even concrete significantly more susceptible to thixotropic dissipation still retains an increased stiffness during the Slump-flow test.

The MPM framework with PR-Bingham model has been successful in offering insights into the flow behaviour of thixotropic concrete. The model suggests that the flow of concrete should be impacted by its thixotropic properties on a greater scale than empirical testing seems to suggest, and in line with observations from a rheometer.

\section{Conclusion} 

The objective of this paper was to model thixotropic Tremie Concrete for the first time, using the Material Point Method to explore the challenges of testing concrete and how numerical modelling can provide unique insights into concrete flow behaviour. We have observed the following:
\begin{itemize}
  \item The new MPM PR-Bingham model that combines the Papanastasiou-Bingham model with Roussel's simplified thixotropy equations is successful in accurately representing concrete flow behaviour in both dynamic and thixotropic conditions. 
  \item MPM simulations of weakly incompressible materials such as tremie concrete results in numerical errors such as volume locking. Volume locking in weakly compressible flows are reduced using higher order shape functions, setting the $m$ parameter in the PR-Bingham model to a low number and using reduced integration for pressure calculations.
  \item A comprehensive calibration and material property assessment procedure, outlined in this paper, can be used to verify simulated Tremie Concrete.
  \item Finally, a comparison of physical and numerical results of the Slump-Flow test for thixotropic concrete reveal for the first time, a decline in concrete workability during the Slump-Flow test for simulated concrete, in line with theoretical predictions, that was not captured by the physical version of the test.
\end{itemize}

\section{Acknowledgements} 

This work was funded through an EPSRC ICASE award (Ref: 1769998) in partnership with industry partner ARUP. The authors thank the CB-GEO computational geomechanics research group for assistance in preparing the simulations. The authors would also like to acknowledge the EFFC and DFI research groups for providing data sets with which to work. 



%%   \citet{<label>} ==> Jones et al. [21]
%%   \citep{<label>} ==> [21]
%%

 \bibliographystyle{elsarticle-num-names} 
 \bibliography{ref.bib}


%\begin{thebibliography}{00}



%\end{thebibliography}
\end{document}